\documentclass[a4paper]{ESASPStyle}
\usepackage{graphicx}

\def\apj{ApJ}%
\def\aap{A\&A}%
\def\aaps{A\&AS}%
\def\mnras{MNRAS}%
\def\solphys{Sol.~Phys.}%
\begin{document}

\title{Simu-LC : a Light-Curve simulator for CoRoT }
\author{F. Baudin\inst{1} 
\and R. Samadi\inst{2}
\and T. Appourchaux\inst{1}  
\and E. Michel\inst{2}}

\institute{Institut d'Astrophysique Spatiale, CNRS/Universit\'e Paris XI UMR 8617,
F-91405 Orsay Cedex, France
\and Laboratoire d'Etudes Spatiales et d'Instrumentation pour
l'Astrophysique, Observatoire de Paris, CNRS UMR 8109, F-92195 Meudon, France}

\maketitle

\begin{abstract}

In order to prepare the analysis of COROT data, it has been decided to build a
simple tool to simulate the expected light curves. This simulation tools takes
into account both instrumental constraints and astrophysical inputs for the
COROT targets. For example, granulation and magnetic activity signatures are
simulated, as well as p modes, with the expected photon noise. However, the
simulations rely sometimes on simple approach of these phenomenons, as the main
goal of this tool is to prepare the analysis in the case of COROT data and not
to perform the most realistic simulations of the different phenomenons.

\end{abstract}

\section{Introduction}

Simulating the data that a space instrument like COROT will provide might
look presomptuous. Indeed, it is certainly, when comparing to previous
comparable instruments like IPHIR or GOLF. These two examples show that
the nominal behaviour of the instrument is not always reached, but this
does not prevent this instrument to provide very interesting data. However,
despite some technical problems, IPHIR and GOLF yielded a wealth of
scientific results. Thus, what is the interest of simulating COROT data?
How close to reality these simualtions will get? This might not be the
most important fact as the preparation of these simulations will help us
to prepare the analysis of real data and to be ready in case of unexpected
technical behaviour of the instrument perturbating the data, or unexpected
physical behaviour of the targets of the instrument. A consequence of that
is that the simulation tool must include technical and physical aspects,
making the task even more difficult. These aspects cover: photon noise,
p modes excitation, granulation signal, stellar activity signal,
orbital perturbations, stellar rotation...

The software presented here is freely available at:\\
{\tt www.lesia.obspm.fr/$\sim$corotswg/simulightcurve.html}

\section{Photon noise}

The photon noise is certainly the easiest component of the noise to simulate...
as far as it has the expected behaviour: a true white noise with a variance
depending on the photon counts. The COROT specifications impose a level of
photon noise of $B_0=0.6$~ppm in the amplitude spectrum of a star of magitude
$m_0=5.7$ for an observing duration of 5 days. The stellar flux for a given
magnitude $m$ being related to
the stellar magnitude by:
\begin{equation}
F=F_0 \, 10^{(m_0 - m)/2.5}
\end{equation}
and knowing that the level of noise varies as the square root of the flux, the
level of noise is related to the star magnitude by:
\begin{equation}
B=B_0 \, 10^{(m_0 - m)/5}.
\end{equation}
As indicated above, this simple relation can model the photon noise but not
some other photon counting perturbations as slow (periods of hours or more)
trends in photon countings which might contributes to the low frequency spectrum
of the noise.

\section{Solar-like oscillations}
\label{sect:solar-like-exc}

The solar-like oscillations are stochastically excited and simulated
here following the recipe of \cite{Anderson90}, recalled below.
Each solar-like oscillation is a superposition of a large number of
excited and damped proper modes. Each solar-like oscillation can then
be decomposed as:
\begin{equation}
\sum_j \, A_j \, exp[-2 i \pi \,  \nu_0 t] \, exp[-\eta (t-t_j)]
H(t-t_j) + {\rm c.c}
\label{simu:eq1}
\end{equation}
where $A_j$ is the amplitude at which the mode $j$ with 
proper frequency $\nu_0$ is excited by turbulent convection, $t_j$ at
which it is excited, $\eta$ is the (linear) mode damping rate, $H$ is
the Heaviside function and ``c.c.'' means complex conjugate.
The Fourier transform of Eq.\,\ref{simu:eq1} yields the spectrum:
\begin{equation}
f(\nu) \simeq {U  \over 1+ 2i (\nu-\nu_0)/\Gamma}
\label{simu:eq2}
\end{equation}
whith  $\Gamma = \eta/\pi$,  $U \equiv \sum_j \tilde A_j$ and $ \tilde A_j$
is a complex number proportional to $A_j \, exp[i \, t_j] $.

As the excitations are random, $t_j$ is random and hence $ \tilde A_j$
has a random phase. As the excitation are very numerous, according to
the central limit princip the real and imaginary parts of the complex
number $U$ are distributed according to a normal statitics.
Hence the spectrum of the oscillations, $f(\nu)$, can be simulated by
generating the real and imaginary parts of the complex
number $U$   according to a normal statitics.
An  inverse  Fourier transform is next applied in order to simulate the
oscillations in the time domain. 
This is the principe of the  \cite{Anderson90}'s recipe. 

The question is now that is the constraints on the mean and variance
of $U$.
From Eq.\,\ref{simu:eq2} we deduce the power spectrum $P(\nu)$ of the
oscillation :
\begin{equation}
P(\nu) = { \| U  \|^2   \over  1+  ( 2 (\nu-\nu_0)/\Gamma) ^2 }
\label{simu:eq3}
\end{equation}
The mean mode profile $<P>$ is obtained by averaging
Eq.\,\ref{simu:eq3} over a larger number of realizations:
\begin{equation}
 < P(\nu) > =  { H  \over  1+  ( 2
  (\nu-\nu_0)/\Gamma) ^2 }
\label{simu:eq4}
\end{equation}
where $H \equiv < \| U  \|^2 >$, which is by definition the
variance of the complex number $U$ since $< U > = 0$. 
We see from Eq.\,\ref{simu:eq4} that the mean profile has a Lorentzian
shape as it  is --\, in first approximation \,-- observed. 

According to the Parseval-Plancherel relation, $H$ is related to the
mean square of the mode intrinsic amplitude in terms  of  luminosity
fluctuations, $A_L^2$, as  (see eg. \cite{Baudin05}): 
\begin{equation}
H  =  { C_\ell^2 \, A_L^2 \over   \pi  \,   \Gamma }
\label{simu:eq5}
\end{equation}
where $C_\ell$ is a visibilty factors, which depends on the degree $\ell$.
Once the mode intrinsic amplitude $A_L$, the
mode line width $\Gamma$ and $C_\ell$ are given, we  have then a
constraint on the variance of $U$. 

Note that the derivation of Eqs.\,\ref{simu:eq2}-\ref{simu:eq5}
assumes that the the mode life-time ($\sim \eta^{-1}$) is much shorter
than the duration of the observation ($T_0$), that is  $T_0\,\eta \gg
1$. For life-times much  longer than $T_0$, different expressions are
derived but the principe of the simulation remains the same.

For sake of simplicity, we use the adiabatic assumption formulated by
 \cite{Kjeldsen95} to deduce the maximum of $A_L$  
from the maximum of the root mean square mode velocity $V_{\rm max}$
 according to: 
\begin{equation}
\left(A_L\right)_{\rm max}= 
\left(A_L \right)_{\odot, {\rm max}}
\frac{V_{\rm max}}{V_{\odot,\rm max}}\sqrt{
\frac{T_{\odot,{\rm eff}}}{T_{\rm eff}}}
\label{conversion}
\end{equation}
where $T_{\rm eff}$ is the effective temperature and the symbol $\odot$ refers to quantities related to the Sun. 
We take for the Sun the \emph{rms values} $(A_L)_{\odot, \rm max}
\simeq 4$~ppm (see table 2 in \cite{Kjeldsen95}
and more recently \cite{Barban04}) and $V_{\odot, \rm max} \simeq $ 27~cm/s 
according to \cite{Chaplin98}'s seismic observations.

In turn the root mean square of the mode intrinsic velocity, $V $,
is related to the rate $\cal P$ at which energy is injected into the
mode by turbulent convection and the mode damping rate $\eta = \pi\, \Gamma$ as:
\begin{equation}
  V^2   = {{\cal P}  \over 2 \, \eta \, {\cal M}  } 
\label{simu:eq6}
\end{equation}
 where ${\cal M} \equiv I / \xi_{\rm r}^2(h) $ is the mode mass, $I$
 the mode inertia, $\xi_{\rm r}$ the mode radial displacement and  $h$
the height above the photosphere where oscillations are measured (we
 consider $h=0$).
The way the quantities involved in Eq.\,\ref{simu:eq6} are model is
 explained in the next two Sects.

\subsection{Mode excitation and damping rates}
 Theoretical mode damping rates are obtained from the tables
 calculated by \cite{Houdek99}. These calculations rely on the
 non-local and time-dependent formulation of
 the convection of \cite{Gough77,Gough76} (for more details, see \cite{Houdek99}).

The computation of the excitation rates $\cal P$ is performed according to the model of
 stochastic excitation of \cite{Samadi00I}.
 The calculations assume -~as in \cite{Samadi02II} ~- a Lorenzian function
 for modelling the convective eddy  time-correlations.
 Furthermore, the characteritic wavenumber $k_0$
 involved in the theory is assumed to be constant according to the simplification proposed by \cite{Samadi02I}.
Its value is related to the  value $k_{0,\odot} \simeq 3.6~ {\rm
 Mm}^{-1}$ inferred in \cite{Samadi02I} from a 3D simulation of the Sun as: 
$k_0 = k_{0,\odot} \, \Lambda_{\odot} / \Lambda$
where $\Lambda$  is the  mixing-length evaluated at the layer the convective
 velocity is maximum.

\subsection{Mode mass and mode  frequency calculation}
\label{Mode mass and mode  frequency calculation}

The solar model we consider is calculated with the CESAM code (\cite{Morel97}) and appropriate input physics, 
described in details in  \cite{Lebreton99}. In particular, convection is modelled according to the classical 
mixing-length theory \cite[hereafter MLT]{Bohm58} with a mixing-length $\Lambda \equiv \alpha H_p$   where $H_p$ is the 
pressure scale height and  $\alpha_c$ is the mixing-length parameter.   
In contrast with   \cite{Lebreton99}, the atmosphere is restored from the Eddington classical gray atmosphere 
and microscopic diffusion is included  according to the simplified formalism of \cite{Michaud93}.
The calibration of the solar model, in luminosity and radius for an age of $4.65$~Gyr, fixes the initial helium 
content $Y=0.2751$, metallicity $Z=0.0196$ and the MLT parameter  $\alpha_c=1.76$.

The oscillation eigenfunctions and hence the mode ma\-sses, $\mathcal{M}$,
in  Eq.~\ref{simu:eq6} are  calculated with the adiabatic ADIPLS pulsation
code (\cite{JCD96}).

\begin{figure*}
\centering
\resizebox{\hsize}{!}{
\includegraphics[angle=90]{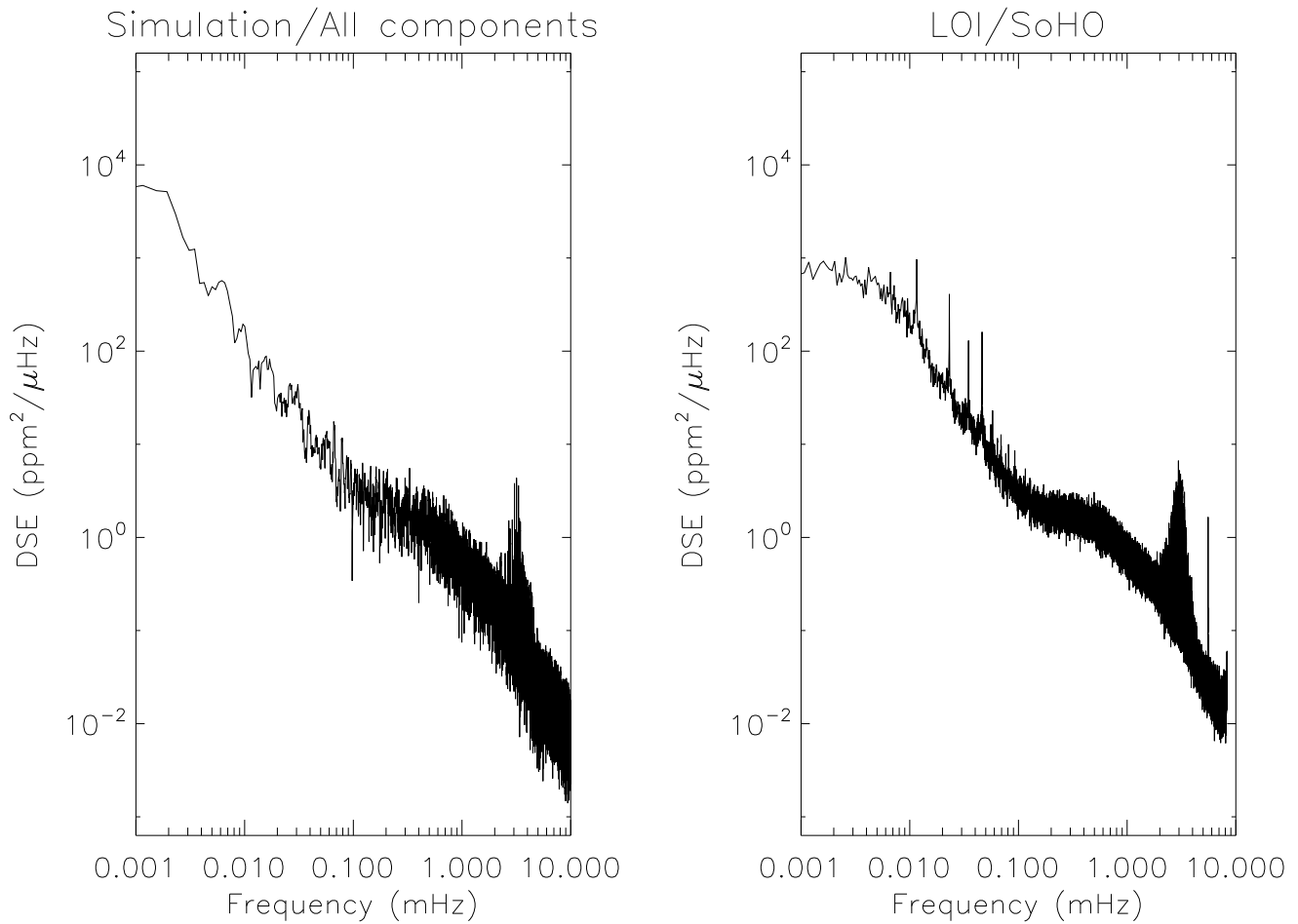}}
\caption{Left: Fourier spectrum of the output of the light curve simulator
described here, in the case of the Sun; Right: Fourier spectra of observation
of the Sun obtained with the photometer LOI onboard SoHO (the high peaks at low
frequency are daily aliases due to SoHO synoptic observations performed every 24
hours). This comparison shows
a reasonnably good agreement, except maybe for the activity signature}
\label{comp_sun}
\end{figure*}

\section{Stellar granulation signal}

Granulation can be considered as noise if the aim of the observation is the
stellar oscillations, but it carries some information about the physics of the
star, and very valuable information as the convection is a very badly described
phenomenon in stellar modelling. Thus, we call it a $signal$ and not a noise.

However, granulation can be described only with a statistical approach.
Moreover, its contribution in the Fourier domain is not independent of
frequency: it will contribute more at low frequency. A common description
is to consider that signal $S$ as a random signal with some memory: then, its
autocorrelation function is:
\begin{equation}
ACF_S(t)=A^2 e^{-|t|/\tau}
\end{equation}
where $A$ is the amplitude of this signal (to be related to its variance) and
$\tau$ a characteristic time. Knowing that the Fourier transform of the
autocorrelation function is the squared modulus of the function $S$, one has:
\begin{equation}
\|S(\nu)\|=\frac{2A^2 \tau}{1 + (2\pi \tau \nu)^2} .
\end{equation}
Thus, the Fourier spectrum of the granulation is modelled as a Lorentzian
which, if it is summed over all frequencies, yields the variance of the signal
and then its relation with $A$:
\begin{equation}
\sigma^2 = \int \|S(\nu)\| \,\, \Rightarrow \,\, \sigma=A/\sqrt{2}.
\end{equation}
So, knowing the intensity standard deviation due to granulation and its characteristic
time, it is possible to model its Fourier spectrum (and considering its phase
is random, it is possible to model the corresponding time series by an
inverse Fourier transform). This is in fact the approach developped by
\cite{Harvey} to model the solar noise spectrum, including granulation,
mesogranulation and supergranulation. It can be used to describe any noise with
memory (some electronic noise for example).

In the case of granulation, the required parameters are estimated from theory of
convection.
The eddy size $\Lambda_{\rm eddy}$ is assumed  to be equal to $ \Lambda = \alpha\, H_p$
where $\alpha$ is the mixing-length parameter and $H_p$ the
pressure scale height.
The eddy overturn time is related to the eddy convective velocity, $v$,
by $\tau_{\rm eddy} = \Lambda_{\rm eddy} / v$ and is considered as the
characteristic time of granulation $\tau$.
The number of number of eddies seen at the star surface,
$N_{\rm eddy}$, is 
$2 \, \pi (R_{*}/\Lambda_{\rm eddy})^2$ where $R_{*}$ is the star radius. 
This relation ignores of course that the medium is higly anisotropic. 
According to the mixing-length theory (see eg. \cite{Cox68}),  the
eddies contrast (border/center of the granule), $(\delta L/L)_{\rm eddy}$,
can be related  to the difference between the temperature
  gradient of the eddy, $\nabla^\prime$ and that of the surrounding
  medium, $\nabla$, according to the relation:
\begin{equation}
(\delta L/L)_{\rm
  eddy} =  {  \nabla  - \nabla^\prime  \over \nabla }
\label{dL_L_eddy}
\end{equation}
In turn, Eq.~\ref{dL_L_eddy} can be reduced to:
\begin{equation}
(\delta L/L)_{\rm
  eddy} =   (4/9) \,  { \lambda \over (1 - \lambda ) } \, {1 \over  \gamma}
\end{equation}
where $\gamma$ is the convection efficiency and $\lambda$ the ratio
between the convective flux and the total flux of energy. 
This relation is finally calibrated in order to match the solar
constraints, and the intensity standard deviation for the whole observed disc
is
\begin{equation}
\sigma = (\delta L/L)_{\rm eddy}/\sqrt{N_{\rm eddy}}
\end{equation}
All the quantities are obtained on 1D stellar models computed as
explained in Sect. \ref{sect:solar-like-exc}.

\section{Stellar activity signal}

The stellar magnetic activity will induce intensity variation in time, mainly
due to the
presence of starspots crossing the observed disc because of the rotation.
However, some
other sources of intensity variations are expected (flares for example).

A first approach is to consider the intensity variations as described by the
same way than granulation. Knowing the standard deviation and the characteristic
time of intensity variations allows to build the Fourier spectrum of these
variations as a Lorentzian. This approach has been used for example by
\cite{Aigrain04}. The difficulty in the case of magnetic activity is that
there is no theoretical description of the phenomenon. Thus, the parameters
describing it are empirically derived.\\
The characteristic time is taken as the period of
rotation of the star, or, if the rotation is slow, the instrinsic lifetime of a
spot. The latter is arbitrarily chosen as the one in the solar case,
computed by \cite{Aigrain04}. The rotation period is computed from an
empirical law involving the age and the $B - V$ color index of the star
described in the same reference.\\
The standard deviation of intensity variations due to magnetic activity is
also derived from an empirical law. However, this law involves the Rossby
number $R_0$ (ratio of rotation period $P_{rot}$ and the overturn convective
time at the base of the convection zone $\tau_{bcz}$). This number can be
derived empirically \cite{Aigrain04},
but in the present simulations, we derive $\tau_{bcz}$ from models. Then we use
empirical laws to estimate $\sigma$.

Another approach is to simulate in the intensity time series the influence of
individual spots, to estimate their number and contrast (as for example
\cite{Lanza04}). The expected result in the Fourier spectrum should be similar
to the first approach, but this detailed simulation should allow for rotation
measurements. This approach will be included in a further version of our
simulation software.

\begin{figure}
\centering
\includegraphics[width=9.5cm]{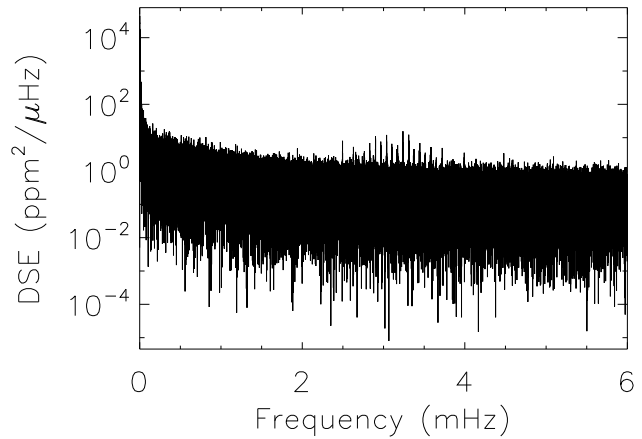}
\caption{Fourier spectra of the output of the simulator described here, in the
case of a Sun seen with a magnitude $m=6$, showing that the p modes are easily
detected, as well as granulation and activity}
\label{sun_m6}
\end{figure}

\section{Examples}

\begin{figure}
\centering
\includegraphics[width=9.5cm]{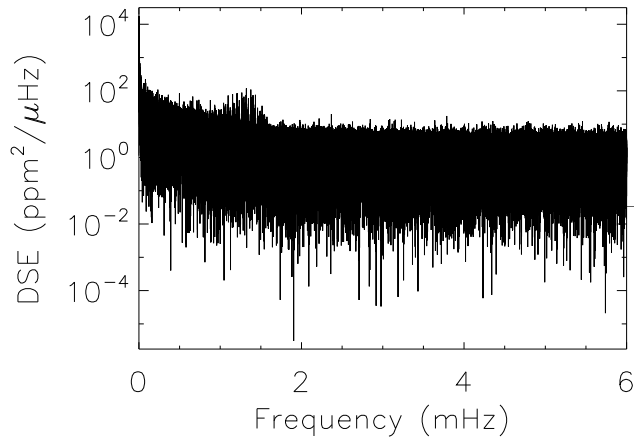}
\caption{Same than Fig.~\ref{sun_m6} for a star with a mass $M=1.5M_{\odot}$ and
a magnitude $m=8$, showing also p modes, granulation and activity signatures}
\label{star1.5_m8}
\end{figure}

A first example of the output of this software is shown in Fig.\ref{comp_sun}: a
simulation of a Sun with a magnitude $m=0$ (in order to have a very weak phoon noise)
is compared with a spectrum from the LOI
photometer (\cite{LOI}) onboard SoHO. The agreement between simulation and
ovservation is good, except maybe at very low frequency: the activity component
of the signal is overestimated in the simulation (this is explained as the Sun
does not fit very well the empirical law used for activity estimation).\\
The following example is a sun with a magnitude $m=6$, showing that both p modes
and activity should be visible with COROT for such a star (Fig.\ref{sun_m6}).
Another example is shown in Fig.\ref{star1.5_m8} for a star of 1.5~solar mass and
an age of 2.4~Gyr: again, p modes and activity are detectable.

\section{Conclusion}

This simulator software will continue to evolve with time. As indicated above,
intensity modulation due to starspots will be included, as well as other stellar
or instrumental signals, as for example instrumental perturbations due to
orbital vraiations.
Moreover, this effort of simulation will not end with the delivery of first data but
will be continued after that. The comparison with real data will allow to
check for the validity of physical hypothesis used to simulate the different
signals of astrophysics origin in the data. This shoud bring a great amount
of information on our knowmedge of these often not well known phenomena,
which stellar simulation is often derived from the solar case.
In parallel, the simulation of instrumental components of the signal
will be improved to help the interpretation of real data. All these
reasons justify in our opinion the need for the simulation tool
presented here.

\begin{acknowledgements}
RS thanks J. Ballot for his help in the computation of stellar models.
\end{acknowledgements}

\bibliographystyle{aa}

\end{document}